\documentclass[aps,prl,floatfix,showpacs,groupedaddress,notitlepage,twocolumn,
amsmath,amssymb,10pt]{revtex4-1}

\usepackage{graphicx}
\usepackage{bm}
\usepackage{subfigure}
\usepackage{braket}
\usepackage{times} 
\usepackage{units} 
  
\usepackage{color}
\definecolor{darkblue}{rgb}{0.2,0.2,0.8}
\usepackage[pdftex,
        colorlinks=true,
        linkcolor=darkblue,
        citecolor=darkblue,
        filecolor=black,
        urlcolor=darkblue,
        bookmarks=false,
        bookmarksopen=false,
        bookmarksopenlevel=3,
        plainpages=false,
        pdfpagelabels=true,
        verbose=true]{hyperref}
\usepackage[all]{hypcap}

\newcommand{\expect}[1]{\langle #1 \rangle}

\newcommand{\Fig}[1]{Fig.~\ref{#1}}

 \newcommand{\Ref}[1]{Ref.~\cite{#1}}

\begin{document}

\title{Notes on the Cluster Gutzwiller Method: \\[0.4ex]Inhomogeneous Lattices, Excitations, and Cluster Time Evolution} 

\author{Dirk-S\"oren L\"uhmann}

\affiliation{\vspace{1ex}Institut f\"ur Laserphysik, Universit\"at Hamburg, Luruper Chaussee 149, 22761 Hamburg, Germany}

\textheight230mm
 

\begin{abstract}
Several perspectives of the cluster Gutzwiller method are briefly discussed. I show that the cluster mean-field method can be used for large inhomogeneous lattices, for computing local excitations, and for the time evolution of correlated quantum systems. 
\end{abstract}

 
\maketitle

Cluster methods have been used for various problems \cite{Bethe1935,Peierls1936,Weiss1948,Senechal2000,Senechal2002,Potthoff2003,Dantziger2002,Du2003,Etxebarria2004,Neto2006,Yamamoto2009,Kotliar2001,McIntosh2012,Pisarski2011,Buonsante2004,Jain2004,Yamamoto2012,Yamamoto2012b} which are typically associated with particles in lattices. The generalization of the widely used Gutzwiller approach \cite{Rokhsar1991,Krauth1992,Zwerger2003,Buonsante2008,Trefzger2011} as a cluster method is described in \Ref{Luhmann2013}. The Gutzwiller method is often understood as an optimization of a trial wave function which is expanded in terms of local Fock states. However, it can also be seen as a direct factorization of an individual lattice site, where the lattice site is represented in a Fock basis. The Gutzwiller approach is suited for the description of weakly interacting bosonic particles and also for the treatment of strongly correlated phases such as the Mott insulator \cite{Zwerger2003}.  

In the standard Gutzwiller method, which is an effective single-site description, it is assumed that we can write the wave function of the system as a product of single-site wave functions. Each site $j$ is represented by the Gutzwiller trial wave function $\ket{j}=\sum_n c_n \ket{n}$ in the basis of local Fock states $\ket{n}$ with $n$ particles \cite{Rokhsar1991,Krauth1992,Zwerger2003,Buonsante2008,Trefzger2011}. Usually the coefficients $c_n$ are determined by imaginary time evolution \cite{Trefzger2011} but the problem can also be solved using a self-consistent diagonalization scheme \cite{Luhmann2013} as applied below.

In the cluster Gutzwiller method \cite{Luhmann2013}, the single-site state $\ket{j}$ is replaced by a cluster $\ket{S}$ with $s$ sites. The factorization approach requires that the wave function is written as a product of cluster and environment wave function. A different mean-field approach for decoupling the cluster is neglecting the fluctuations of quadratic order \cite{Oosten2001,McIntosh2012,Pisarski2011}. For the standard Bose--Hubbard model both approaches are equivalent and lead to the same results \cite{Krauth1992,Sheshadri1993}. The internal degrees of freedom within the cluster allow for quantum fluctuations that are not covered within the single-site mean-field approach. The generalized trial wave function of the cluster is given in a multi-site Fock basis $\{\ket{N}\}=\{\ket{n_0,n_1,n_2,...}\}$ accounting for all possible distributions of $n_i=0,1,2,...$ particles on the cluster sites $i$. Using the factorization, the wave function of the whole system can be written as $\ket{S}\ket{\psi}$, where $\ket{\psi}$ is the wave function describing the state outside the cluster. 

Common tight-binding Hamiltonians can be written as 
\begin{equation} 
	\hat H=\hat H_\psi + \hat H_S + \hat H_{\psi S}  =\hat H_\psi + \hat H_S + {\sum}_\alpha \hat A_{\psi}^\alpha\, \hat B_{S}^\alpha ,
\end{equation}
where $\hat H_\psi$ and $\hat H_S$ act only on the respective subsystems.  
The last term represents the coupling between $\ket{\psi}$ and $\ket{S}$ and can be decomposed in a sum of subsystem operator products. The Hamiltonian matrix of the whole system in the cluster basis $\{\ket{N}\}$ is given by
\begin{equation} \begin{split}
	H_{MN}=&\bra{\psi} \hat H_\psi \ket{\psi} \delta_{MN} + \bra{M} \hat H_j \ket{N} +\\
	& {\sum}_\alpha \bra{\psi} \hat A_\psi^\alpha \ket{\psi}    \bra{M} \hat B_j^\alpha \ket{N},
	\label{Eq2}
\end{split}\end{equation}
where the first term is a constant energy offset. In this factorized expression, the Hamiltonian depends only on expectation values of $\ket{\psi}$. These expectation values naturally define the mean-field parameters and are
determined in a self-consistent loop. Usually, the mean fields are expectation values of a cluster site. For the standard Bose--Hubbard model and a homogeneous lattice, the last term in Eq.~\eqref{Eq2} reduces to $-J \sum_{j\in \partial S} \bra{M} \nu_j \Big( \hat a^\dagger_{j} \expect{\hat a}  + \hat a_{j}  \expect{ \hat a^\dagger} \Big)  \ket{N}$ and describes the coupling of all sites at the boundary $\partial S$ of the cluster with the mean-field $\expect{\hat a}$. The prefactor $\nu_j$ reflects the number of bonds to the mean-field.

\begin{figure}[b]
\centering 
\includegraphics[width=0.65\linewidth]{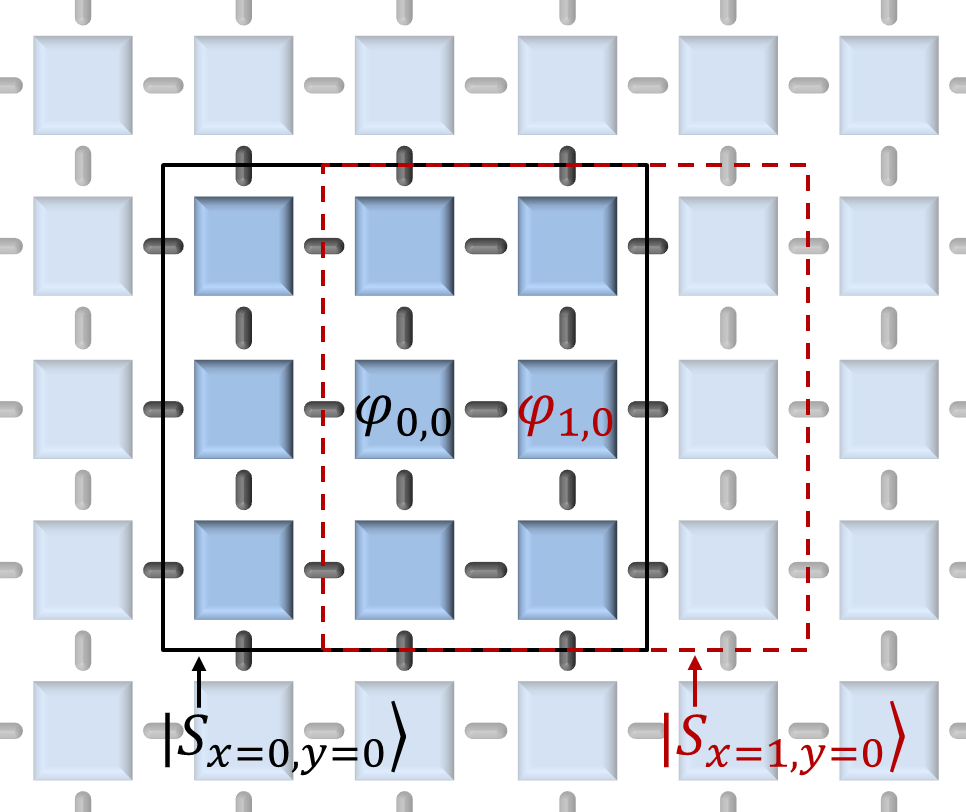} 
\caption{An inhomogeneous lattice can be covered with overlapping clusters $\ket{S_{x,y}}$. In each self-consistent iteration the ground state of each cluster and the mean-field parameters on the target sites are determined. After several iterations cluster ground states and mean-field parameters converge. The same technique can be used for computing the time evolution of an initial state [see Eq.~\eqref{Eq4}].
}
\label{Figure1}
\end{figure}

\begin{figure*}[t]
\centering 
\includegraphics[width=0.8\linewidth]{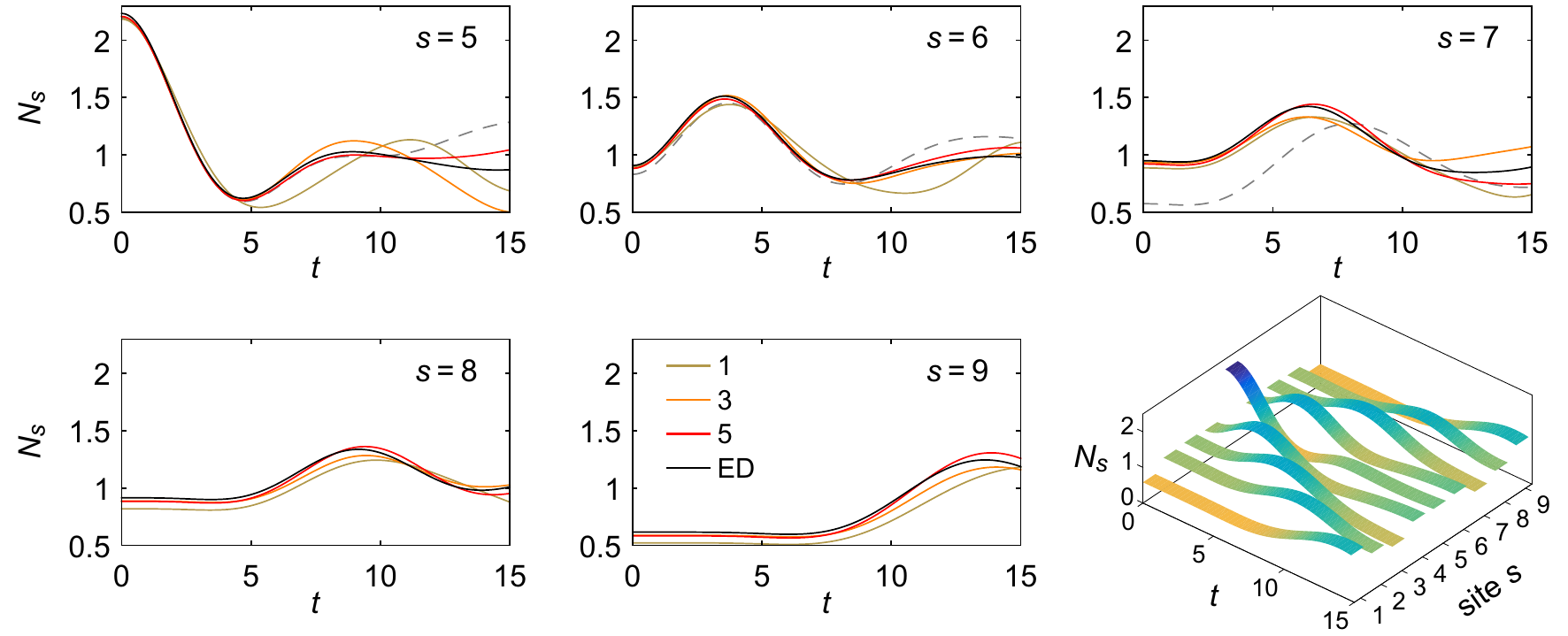} 
\caption{Time propagation of a particle excitation in the Bose--Hubbard model for a one-dimensional lattice with nine sites and hard boundaries. The initial state is the ground state for the tunneling amplitude $J=0.5 U$, the on-site interaction $U$, and one particle per site (the chemical potential is adapted). At time $t=0$, an additional particle is created on the central site $s=5$. The particle densities $N_s$ are shown for the sites $s=5,...,9$ using clusters with one, three, and five sites and are compared with the exact solution for nine sites (black line). For comparison the dashed gray line depicts the exact solution for five sites. The last plot shows the results of the cluster time evolution using five-site clusters. The time $t$ is given in units of $hU/J$. 
}
\label{Figure2}
\end{figure*}

The wave function of the cluster can be obtained by solving the eigenvalue problem $H_{MN}$ of the cluster. In turn, the lowest eigenvector $\ket{S}= \sum_N C_N \ket{N}$ is used for computing the mean-field parameters on a target site $t$, e.g., 
\begin{equation}
	\expect{\hat a_t} =\bra{S} \hat a_{t} \ket{S} = \sum_{M,N} C_M^* C_N \bra{M} \hat a_{t} \ket{N}.
\end{equation}
Thus, using a self-consistent loop, the Hamiltonian can be solved without initial knowledge of $\ket{\psi}$. For more complicated Hamiltonians with off-site interactions or occupation-dependent processes \cite{Dutta2015}, $\bra{\psi} \hat A_{\psi}^\alpha \ket{\psi} $ is not reducible to a single mean field $\expect{\hat a_i}$ but can contain several other mean-field parameters. This is demonstrated in \Ref{Jurgensen2015}, where an extended Hubbard model is considered giving rise to three and eight mean-field parameters for one and two components, respectively. These parameters are a direct consequence of the applied factorization and take the role of order parameters, which are used to define the phase boundaries. 

For the accuracy of the cluster method the ratio of internal bonds in the cluster to mean-field bonds at the boundary is crucial. By applying a scaling of the cluster size, the critical $J/U$ for an infinite lattice can be interpolated \cite{Yamamoto2012b,Luhmann2013}. The scaling parameter  reads $\lambda=B_S / (B_S+B_{\partial S})$, where $B_S$ represents the number of bonds within the cluster and  $B_{\partial S}$ the bonds to the mean fields. The scaling parameter also explains why periodic boundary conditions \cite{Luhmann2013} improve the numerical results significantly. 

In the following, it is shown that the cluster Gutzwiller method can also be used for large inhomogeneous lattices, for computing local excitations, and for performing time evolution. In \Ref{Jurgensen2014} the cluster method was applied to lattices with inequivalent lattice sites in confined systems. In general, it can be used for spatial-varying lattices, large unit cells, or disorder potentials \cite{Pisarski2011}. A given lattice can be covered with (overlapping) clusters centered at lattice sites $j$ as shown in \Fig{Figure1}. The self-consistent loop must now performed for a set of clusters, where the order within each iteration is usually not important. In each iteration the ground state of each cluster is computed and the the mean-field parameters, e.g.~the superfluid order parameter $\varphi_{r}= \expect{\hat a_{r}}$, at target sites $r$ are updated until the mean fields and the eigenstates of the clusters converge. Furthermore, the cluster mean-field method allows us to compute local excitations within the clusters, where typically only the low-lying eigenstates are of interest.  The excitations correspond to the eigenstates of the clusters where the mean-field boundary is determined from the ground state. Such a local eigenspectrum was computed for a dimerized Mott insulator phase in a honeycomb lattice in \Ref{Jurgensen2014}. 

Dynamical problems can also be solved with the cluster Gutzwiller method. While the time-dependent DMRG method can be used for one-dimensional systems \cite{Daley2004}, the time evolution for higher dimensions is usually limited to either small systems applying exact methods or to uncorrelated mean-field techniques \cite{Jaksch2002,Damski2003,Trefzger2011,Bissbort2011,Fischer2011}. The cluster time evolution accounts for short-range correlations in an exact way, whereas long-range correlations are modeled by time-dependent mean-field parameters. As described above, a lattice can be composed of overlapping clusters. For clusters with an odd number of sites in each dimension, each lattice site $r$ can be associated with a cluster centered at this site (except for edge sites that are non-centered target sites of edge clusters). Thus, in the bulk of the lattice we have a set of mean-field parameters (e.g.~the superfluid order parameter $\varphi_{r}$) and an associated cluster for each lattice site. The key for performing correlated time evolution is that the cluster state $\ket{\psi^r_t}= \sum_N p^{r}_N(t) \ket{N}$ is known at time $t$, where $\ket{N}$ is the many-particle cluster basis. For instance, we can start with the ground state of the system assuming that the Hamiltonian is only time-dependent for $t>t_0$. In this case, the state $\ket{\psi^r_t}= \ket{\phi^r_0}$ is the lowest eigenvector of the cluster $r$, where the stationary mean-field parameters at its boundary are computed self-consistently from the neighboring clusters for $t<t_0$.

At time $t_0$, either the Hamiltonian becomes time-dependent (e.g.~because of a quench or because the system is externally driven) or by a process that adds or removes particles. At any time $t>t_0$, we can diagonalize the Hamiltonian $H^r_{t}$ of the cluster $r$ in dependence on the mean-field parameters of the neighboring sites evaluated at time $t$. The result is the complete eigenbasis of the cluster $\ket{\phi^r_n}=\sum_N c_N^n(t) \ket{N}$ with eigenenergies $\hbar \omega_n$. After a short time step $\Delta t$ at the time $t'=t+\Delta t$, the cluster wave function reads 
\begin{equation}\begin{split}
\ket{\psi^r_{ t+\Delta t}}&=\sum_n e^{-i \omega_n \Delta t} \braket{\phi^r_n|\psi^r_t} \ket{\phi^r_n}\\
&= \sum_N \ket{N} \overbrace{\sum_n c_N^n(t) e^{-i \omega_n \Delta t} \sum_M c_M^{n*}(t)\, p^r_M(t)}^{p^r _N(t+\Delta t)}
\label{Eq4}
\end{split}\end{equation}
with $\braket{\phi^r_n|\psi^r_t}=\sum_ M c_M^{n*}(t)\, p^r_M(t)$. From the coefficients $p^r_N(t+\Delta t)$ we can calculate the mean-field expectation values at time $t+\Delta t$ (e.g. $\varphi_r(t+\Delta t)$). 
For a better visualization, the example in \Fig{Figure2} depicts the cluster time evolution of a one-dimensional problem. However, the main application for this method is performing the time propagation using  Eq.~\eqref{Eq4} for two- and three-dimensional lattices.

In conclusion, the cluster Gutzwiller method allows the computation of accurate phase boundaries for infinite lattices, of finite inhomogeneous systems, of local excitations, as well as the time evolution of correlated quantum systems.   
 

%

\end{document}